\documentclass[aps,showpacs,epsfig]{revtex4}

\usepackage{epsfig}
\usepackage{color}
\usepackage{bbm}
\usepackage{epsfig}
\usepackage{color}
\usepackage{amsmath}
\usepackage{amssymb}
\usepackage{bbm}

\newcommand{\be}{\begin{equation}}
\newcommand{\ee}{\end{equation}}
\newcommand{\bea}{\begin{eqnarray}}
\newcommand{\eea}{\end{eqnarray}}
\usepackage{color}
\begin{document}

\title{\bf\Large {Electrodynamics of s-wave superconductors}}

\author{Naoum Karchev\footnote{Tel: +3 592 8527754 \\ E-mail address: naoum@phys.uni-sofia.bg } }

\affiliation{Department of Physics, University of Sofia, 1126 Sofia, Bulgaria }

\begin{abstract}

In this paper we give a derivation of a system of equations to describe the electrodynamics of s-wave superconductors.
First, we consider a relativistically covariant theory in terms of gauge four-vector electromagnetic potential and scalar complex field. We use
the first-order formalism to obtain the supplemented Maxwell equations for gauge invariant electric, magnetic, four-vector fields
and the modulus of the superconducting order parameter. The new four-vector field appears in some of the equations as a gauge invariant super-current and in other ones, while gauge invariant, as a four-vector electromagnetic potential. This dual contribution of the new four-vector field is the basis of the electrodynamics of superconductors.  We focus on the system of equations with time-independent fields. The qualitative analysis shows that the applied magnetic field suppresses the superconductivity, while the applied electric field impacts appositely, supporting it. Second we consider time-dependent non-relativistic Ginzburg-Landau theory.

\end{abstract}


\maketitle

\section {\bf Introduction}

The earliest study of the electrodynamics of s-wave superconductors is attributed to London brothers \cite{London35}. They supplemented the Maxwell system of equations with set of equations to explain the electrodynamics of superconductors and more particularly the Meissner-Ochsenfeld effect. The quantum-mechanical foundation of these equations was discussed in phenomenological Ginzburg-Landau theory \cite{GL50}. There is also an attempt to explain the Meissner-Ochsenfeld effect in a purely classical way \cite{Edwards81}.

The unsuccessful experiments to detect an electric field in superconductors \cite{London36} led to the lack of interest. There is at present no general understanding of the interplay between applied electric, magnetic fields and superconductivity. In \cite{Hirsch03} a microscopic justification is given  that a superconductor may have an electric field in its interior. The phenomenon is considered as a consequence of hole superconductivity \cite{Hirsch89}.

Our main goal is to show that a system of equations which describes the electrodynamics of s-wave superconductors can be derived from time dependent Ginzburg-Landau theory. First, we consider a relativistically covariant theory in terms of gauge four-vector electromagnetic potentials and scalar complex field. The electrodynamics is a Lorentz covariant theory and one expects that the model under consideration will help to get deeper insight for the interplay between electric, magnetic fields and superconductivity. We want also to compare our results with the results in \cite{London35} and \cite{Hirsch04} where relativistically covariant theory of superconductivity is discussed. We focus on the system of equations with time-independent fields. The qualitative analysis shows that the applied magnetic field suppresses the superconductivity, while the applied electric field impacts appositely, supporting it.

The system of equations, derived from a relativistically non-covariant theory, shows that the effect of the applied electric field depends on the direction of the field.

The paper is organized as follows: In Section  II, we derive the system of equations to describe the electrodynamics of s-wave superconductors from relativistically covariant theory of superconductivity. For the case when density of Cooper pairs is a constant the system of equations reduces to the London brothers equations. The system of equations derived in the present paper includes an equation for the density of Cooper pairs which shows the different impact on superconductivity of applied electric and magnetic fields.  In Section III, we use the same technique of calculations to consider time-dependent relativistically non-covariant Ginzburg-Landau theory. The main results are reported and commented in Section IV.

\section {\bf Relativistically covariant theory of superconductivity}

We begin with a well-known field-theory action \cite{Higgs64} for relativistically covariant theory of superconductivity
\bea\label{MSc1}
S & = & \int d^4x\left [-\frac 14 \left (\partial_{\lambda}A_{\nu}-\partial_{\nu}A_{\lambda}\right)\left (\partial^{\lambda}A^{\nu}-\partial^{\nu}A^{\lambda}\right)\right. \nonumber\\
& + & \left.  \left (\partial_{\lambda}-ie^*A_{\lambda}\right)\psi^* \left (\partial^{\lambda}+ie^*A^{\lambda}\right)\psi \right. \\
& + & \left. \alpha \psi^*\psi -\frac {g}{2}\left(\psi^*\psi \right)^2\ \nonumber \right]
\eea
written in terms of gauge four-vector electromagnetic potential $"A"$ and complex scalar field $"\psi"$, the superconducting order parameter. The parameter
\be\label{MSc12}
\alpha=\alpha_0(T_c-T,)\ee where $T$ is the temperature and $T_c$ is the critical temperature, is positive when the system is superconductor. We use the standard, for the relativistically covariant systems, notations: $x=(x^0,x^1,x^2,x^3)=(x_0,-x_1,-x_2,-x_3)=(\upsilon t,x,y,z)$, $\upsilon^{-2}=\mu\varepsilon$, where $\mu$ is the magnetic permeability and $\varepsilon$ is the electric permittivity of the superconductor. We assume that these parameters do not change their values when the system undergoes normal to superconductor transition. The action (\ref{MSc1}) is invariant under the gauge transformations
\bea\label{MSc2}
\psi'(x) & = & \exp {[i e^*\phi(x)]}\psi \nonumber \\
A'_{\nu} & = & A_{\nu}-\partial_{\nu}\phi(x),
\eea
where $\phi(x)$ is a real function.

We represent the order parameter $\psi(x)$ in the form
\be\label{MSc3}
\psi(x) =  \rho(x)\exp {[ie^*\theta(x)]}, \ee
where $\rho(x)=|\psi(x)|$ is a gauge invariant and the gauge transformation of $\theta(x)$ is
\be\label{MSc4}
\theta'(x) =  \theta(x)+\phi(x).\ee
The action (\ref{MSc1}), rewritten in terms of $\rho$ and $\theta$, adopts the form
\bea\label{MSc5}
S & = & \int d^4x\left [-\frac 14 \left (\partial_{\lambda}A_{\nu}-\partial_{\nu}A_{\lambda}\right)\left (\partial^{\lambda}A^{\nu}-\partial^{\nu}A^{\lambda}\right)\right. \nonumber\\
& + & \left.  e^{*2}\rho^2\left (\partial_{\lambda}\theta+ A_{\lambda}\right)\left (\partial^{\lambda}\theta+A^{\lambda}\right) \right. \\
& + & \left. \partial_{\lambda}\rho\partial^{\lambda}\rho +  \alpha \rho^2 -\frac {g}{2}\rho^4 \right].\nonumber
\eea

It is convenient to use the action in the first-order formalism
\bea\label{MSc6}
S & = & \int d^4x\left \{-\frac 12 \left [\left (\partial_{\lambda}A_{\nu}-\partial_{\nu}A_{\lambda}\right)F^{\lambda\nu}
-\frac 12 F_{\lambda\nu}F^{\lambda\nu}\right]\right. \nonumber \\
& + & \left.  2e^{*2}\rho^2\left[\left (\partial_{\lambda}\theta+ A_{\lambda}\right)Q^{\lambda}-\frac 12 Q_{\lambda}Q^{\lambda}\right] \right.  \\
& + & \left. \partial_{\lambda}\rho\partial^{\lambda}\rho +  \alpha \rho^2 -\frac {g}{2}\rho^4  \right\},\nonumber
\eea
where gauge potential $A^{\lambda}$, phase $\theta$,  gauge invariant antisymmetric field $F^{\lambda\nu}=-F^{\nu\lambda}$, gauge invariant four-vector field $Q^{\lambda}$ and gauge invariant scalar field $\rho$ are assumed to be independent degrees of freedom in the theory.

To derive the system of Maxwell equations for s-wave superconductors we vary the action (\ref{MSc6}) with respect to $F^{\lambda\nu}$, $Q^{\lambda}$, $A^{\lambda}$, $\theta$ and $\rho$. The resulting system of equation reads:
\bea
& & F_{\lambda\nu}\, = \, \left (\partial_{\lambda}A_{\nu}-\partial_{\nu}A_{\lambda}\right)\label{MSc71} \\
& & Q_{\lambda}\, = \,\partial_{\lambda}\theta+ A_{\lambda}\label{MSc72} \\
& & \partial_{\lambda}F^{\lambda\nu}\, + \, 2e^{*2}\rho^2Q^{\nu} = 0 \label{MSc73} \\
& & \partial_{\lambda}\left(\rho^2Q^{\lambda}\right) \, = \, 0 \label{MSc74}\\
& & \partial_{\lambda}\partial^{\lambda}\rho-\alpha \rho+g\rho^3 \nonumber \\
& & = 2e^{*2}\rho\left[\left (\partial_{\lambda}\theta+ A_{\lambda}\right)Q^{\lambda}-\frac 12 Q_{\lambda}Q^{\lambda}\right]\label{MSc75}
\eea


If we set in equations (\ref{MSc73}), (\ref{MSc74}) and (\ref{MSc75}) the expressions for $ F_{\lambda\nu}$ and  $ Q^{\lambda}$ from equations (\ref{MSc71}) and (\ref{MSc72}) we obtain the equations of motion following from the action (\ref{MSc5}). This means that theories with actions (\ref{MSc5}) and (\ref{MSc6}) are equivalent.

Alternatively, one eliminates the gauge fields $A_{\lambda}$ and $\theta$ from equations (\ref{MSc71}-\ref{MSc75}) to obtain the system of equations for the gauge invariant fields $F^{\lambda\nu}$,  $Q^{\lambda}$ and $\rho$:
\bea
& & \partial_{\lambda}F_{\nu\delta}\,+\,\partial_{\nu}F_{\delta\lambda}\,+\,\partial_{\delta}F_{\lambda\nu}\,=\,0\label{MSc81} \\
& & \partial_{\lambda}F^{\lambda\nu}\,+\,2e^{*2}\rho^2Q^{\nu}\,=\,0 \label{MSc82}\\
& & \partial_{\lambda}\left (\rho^2 Q^{\lambda}\right )\,=\,0 \label{MSc83}\\
& & \partial_{\lambda} Q_{\nu}\,-\,\partial_{\nu} Q_{\lambda}\,=\,F_{\lambda\nu} \label{MSc84}\\
& & \partial_{\lambda}\partial^{\lambda}\rho-\alpha \rho+g\rho^3- e^{*2}\rho Q_{\lambda}Q^{\lambda}=0 \label{MSc85}.
\eea
Equation (\ref{MSc81}) follows from equation (\ref{MSc71}), while equation (\ref{MSc84}) from equation (\ref{MSc72}).

Straightforward calculations show that equation (\ref{MSc83}) can be obtained from equation (\ref{MSc82}) and equation (\ref{MSc81}) from equation (\ref{MSc84}). The system of independent equations is:
\bea
& & \partial_{\lambda}F^{\lambda\nu}\,+\,2e^{*2}\rho^2Q^{\nu}\,=\,0 \label{MSc91}\\
& & \partial_{\lambda} Q_{\nu}\,-\,\partial_{\nu} Q_{\lambda}\,=\,F_{\lambda\nu} \label{MSc92}\\
& & \partial_{\lambda}\partial^{\lambda}\rho-\alpha \rho+g\rho^3- e^{*2}\rho Q_{\lambda}Q^{\lambda}=0 \label{MSc85}.
\eea

We construct the antisymmetric tensor $F_{\lambda\nu}$ by means of the electric $\bf E$ and magnetic $\bf B$ fields in a standard way:
$(F_{01},F_{02},F_{03})=\textbf{E}/\upsilon$, $(F_{32},F_{13},F_{21})=\textbf{B}$ and $(Q^0,Q^1,Q^2,Q^3)=(Q/\upsilon,\textbf{Q})$. In terms of
$\textbf{E},\textbf{B},\textbf{Q}$ and $Q$ the system of equations which describes the electrodynamics of s-wave superconductors is:
\bea
& & \overrightarrow{\nabla}\times\textbf{B}\,=\,\mu\varepsilon\frac {\partial \textbf{E}}{\partial t}-2e^{*2}\rho^2\textbf{Q} \label{MSc101}\\
& & \overrightarrow{\nabla}\times\textbf{Q}\,=\,\textbf{B}\label{MSc102}\\
& & \overrightarrow{\nabla}\cdot\textbf{E}\,=\,-2e^{*2}\rho^2Q \label{MSc103}\\
& & \overrightarrow{\nabla} Q+\frac {\partial \textbf{Q}}{\partial t}\,=\,-\textbf{E}\label{MSc104}\\
& & \mu\varepsilon\frac {\partial^2 \rho}{\partial t^2}-\Delta\rho -\alpha \rho+g\rho^3- e^{*2}\rho\left [\mu\varepsilon Q^2-\textbf{Q}^2\right]=0. \nonumber \label{MSc105}\\
\eea
It is important to stress that the gauge invariant vector $\textbf{Q}$ and scalar $Q$ fields take part in the equations (\ref{MSc102}) and (\ref{MSc104}) as a magnetic vector and electric scalar potentials, while in equation (\ref{MSc101}) $(-2e^{*2}\rho^2\textbf{Q})$ is a supercurrent and in equation (\ref{MSc103}) $(-2e^{*2}\rho^2Q)$ is a density of superconducting quasi-particles . This dual contribution of the new fields is the basis of the electrodynamics of superconductors.

We focus on the system of equations with time-independent fields:
\bea
& & \overrightarrow{\nabla}\times\textbf{B}\,=\,-2e^{*2}\rho^2\textbf{Q} \label{MSc111}\\
& & \overrightarrow{\nabla}\times\textbf{Q}\,=\,\textbf{B}\label{MSc112}\\
& & \overrightarrow{\nabla}\cdot\textbf{E}\,=\,-2e^{*2}\rho^2Q \label{MSc113}\\
& & \overrightarrow{\nabla} Q\,=\,-\textbf{E}\label{MSc114}\\
& & \Delta\rho +\alpha \rho-g\rho^3+ e^{*2}\rho\left [\mu\varepsilon Q^2-\textbf{Q}^2\right]=0.  \label{MSc115}
\eea

To compare our result with the London equations \cite{London35} we assume that deep inside the superconductor $\textbf{Q}$ and $Q$ are zero, while $\rho=\rho_0$ is a constant determined from the equation $\alpha \rho_0-g\rho_0^3=0$, which follows from equation (\ref{MSc115}). The gauge transformation (\ref{MSc4}) of the phase of the order parameter $\theta$ implies that one can impose the gauge fixing condition $\theta=0$. In that case the gauge invariant vector is equal to the vector potential $\textbf{Q}=\textbf{A}$, the gauge invariant scalar field is equal to the scalar potential $Q=A^0$ and the system of equations (\ref{MSc111}-\ref{MSc114}) adopts the form
\bea
& & \overrightarrow{\nabla}\times\textbf{B}\,=\,-2e^{*2}\rho_0^2\textbf{A} \label{MSc111a}\\
& & \overrightarrow{\nabla}\times\textbf{A}\,=\,\textbf{B}\label{MSc112a}\\
& & \overrightarrow{\nabla}\cdot\textbf{E}\,=\,-2e^{*2}\rho_0^2A^0 \label{MSc113a}\\
& & \overrightarrow{\nabla} A^0\,=\,-\textbf{E}\label{MSc114a}.
\eea
With the London brothers postulates in mind
\bea
& & \textbf{J}/c\,=\,-2e^{*2}\rho_0^2\textbf{A} \label{MSc111b}\\
& & \rho\,=\,-2e^{*2}\rho_0^2A^0 \label{MSc113b},
\eea
we arrived at London equations \cite{London35} .

Taking the curl of  (\ref{MSc111a}), using the equation (\ref{MSc112a}) and the identity $ \overrightarrow{\nabla}\cdot\textbf{B}=0$, which follows from this equation, we obtain
\be\label{Msc1201a}
\Delta \textbf{B}\,=\,\frac {1}{\lambda_L^2}\textbf{B},\ee
where 
\be\label{MSc1202}\lambda_L=\sqrt{g/(2e^{*2}\alpha)}.\ee
Taking the divergence of (\ref{MSc114a}) and using the equation (\ref{MSc113a}) we obtain
\be\label{Msc1201b}
\Delta \textbf{E}\,=\,\frac {1}{\lambda_L^2}\textbf{E}.\ee
The equations (\ref{Msc1201a}) and (\ref{Msc1201b}) imply that an electric field penetrates a distance
$\lambda_L$ as a magnetic field does \cite{Hirsch04}.

This approximation is very rough and does not account for the last term in the equation (\ref{MSc115}) which is responsible for the different impact on superconductivity of applied electric and magnetic fields. If we apply magnetic field ($\textbf{E}=0,Q=0$) the qualitative analysis of equation (\ref{MSc115}) shows that the magnetic vector potential effectively decreases the $\alpha$ parameter, $\alpha\rightarrow\alpha-e^{*2}<\textbf{Q}^2>$, where $<\textbf{Q}^2>$ is some average value. Therefor the Ginzburg-Landau coherence length increases \cite{Appendix}, which means that applied magnetic field destroys superconductivity. On the other hand, when the electric field is applied  ($\textbf{B}=0,\textbf{Q}=0)$ the electric scalar potential effectively increases the $\alpha$ parameter $\alpha\rightarrow\alpha+e^{*2}\mu\varepsilon<Q^2>$. Hence, the Ginzburg-Landau coherence length decreases.  This qualitative analysis permits us to formulate the hypotheses that applying electric field at very low temperature  one increases the critical magnetic field. This result is experimentally testable.

\section {\bf Time dependent Ginzburg-Landau theory}

A number of authors have discussed the non-relativistic time-dependent generalization of the Ginzburg-Landau theory \cite{Abrahams66,Caroli67,Gorkov68,Thompson70,Tinkham75}. We investigate a model with field-theory action
\bea\label{MSc14}
S & = & \int d^4x\left [-\frac 14 \left (\partial_{\lambda}A_{\nu}-\partial_{\nu}A_{\lambda}\right)\left (\partial^{\lambda}A^{\nu}-\partial^{\nu}A^{\lambda}\right)\right. \nonumber\\
& + & \left.  \frac {1}{D}\psi^*\left (i\partial_{t}-e^*\varphi\right)\psi \right. \\
& - & \left.  \frac {1}{2m^*} \left (\partial_{k}-ie^*A_{k}\right)\psi^* \left (\partial_{k}+ie^*A_{k}\right)\psi \right. \nonumber \\
& + & \left. \alpha \psi^*\psi -\frac {g}{2}\left(\psi^*\psi \right)^2\ \nonumber \right],
\eea
where $\varphi=\upsilon A_0$ is the electric scalar potential, with gauge transformation (\ref{MSc2}) $\varphi'=\varphi-\partial_{t}\phi$,  $D$ is the normal-state diffusion constant \cite{Tinkham75}, and ($e^*,m^*$) are effective charge and mass of superconducting quasi-particles. The index $k$ runs $k=x,y,z$.

We follow the same procedure to derive the system of equations which describe the electrodynamics of s-wave superconductors. We represent the order parameter $\psi$ by means of modulus and phase (\ref{MSc3}) and write the field-theory action, in the first-order formalism, in the form
\bea\label{MSc15}
S & = & \int d^4x\left \{-\frac 12 \left [\left (\partial_{\lambda}A_{\nu}-\partial_{\nu}A_{\lambda}\right)F^{\lambda\nu}
-\frac 12 F_{\lambda\nu}F^{\lambda\nu}\right]\right. \nonumber \\
& - & \left. \frac {e^*}{D}\rho^2\left (\varphi+\partial_t\theta\right)\right. \\
& - & \left.  \frac {e^{*2}}{m^*}\rho^2\left[\left (\partial_{k}\theta+ A_{k}\right)Q_{k}-\frac 12 Q_{k}Q_{k}\right] \right. \nonumber  \\
& - & \left. \frac {1}{2m^*}\partial_{k}\rho\partial_{k}\rho +  \alpha \rho^2 -\frac {g}{2}\rho^4  \right\}.\nonumber
\eea
It is important to stress that $Q_k$ is a three component vector field and $k=x,y,z$. Using a variational principle we obtain the system of equations
\bea
& & F_{\lambda\nu}\, = \, \left (\partial_{\lambda}A_{\nu}-\partial_{\nu}A_{\lambda}\right)\label{MSc161} \\
& & Q_{k}\, = \,\partial_{k}\theta+ A_{k}\label{MSc162} \\
& & \partial_{\lambda}F^{\lambda}_{\,\,\,\,\,k}\, + \, \frac {e^{*2}}{m^*}\rho^2Q_{k} = 0 \label{MSc163} \\
& & \partial_{k}F_{0k}\, - \, \frac {\upsilon e^{*}}{D}\rho^2 = 0 \label{MSc163} \\
& & \partial_{t}\rho^2+\frac {D}{m^*}\partial_{k}\left(\rho^2Q_{k}\right) \, = \, 0 \label{MSc164}\\
& & \frac{1}{2m*}\Delta\rho+\alpha \rho-g\rho^3 -\frac {e^*}{D}\rho\left(\varphi+\partial_{t}\theta\right)                                      \nonumber \\
& & = \frac {e^{*2}}{m^*}\rho\left[\left (\partial_{k}\theta+ A_{k}\right)Q_{k}-\frac 12 Q_{k}Q_{k}\right].\label{MSc165}
\eea
We supplement the system of equations (\ref{MSc161}-\ref{MSc165}) with equation
\be\label{MSc17}
Q\, = \,\partial_{t}\theta+ \varphi,\ee
which is a definition of the new gauge invariant field $Q$. In the same way, starting from the system of equations (\ref{MSc161}-\ref{MSc17}),  we arrive at the Maxwell equations for superconductors in a non-relativistic theory.
\bea
& & \overrightarrow{\nabla}\times\textbf{B}\,=\,\mu\varepsilon\frac {\partial \textbf{E}}{\partial t}-\frac {e^{*2}}{m^*}\rho^2\textbf{Q} \label{MSc181}\\
& & \overrightarrow{\nabla}\times\textbf{Q}\,=\,\textbf{B}\label{MSc182}\\
& & \overrightarrow{\nabla}\cdot\textbf{E}\,=\,\frac {\mu\varepsilon e^{*}}{D}\rho^2 \label{MSc183}\\
& & \overrightarrow{\nabla} Q+\frac {\partial \textbf{Q}}{\partial t}\,=\,-\textbf{E}\label{MSc184}\\
& & \frac {1}{2m^*}\Delta\rho +\alpha \rho-g\rho^3- \frac {e^{*}}{D} \rho Q-\frac {e^*}{2m^*}\rho \textbf{Q}^2=0. \label{MSc185}
\eea

The system of equations for time-independent fields is:
\bea
& & \overrightarrow{\nabla}\times\textbf{B}\,=\,-\frac {e^{*2}}{m^*}\rho^2\textbf{Q} \label{MSc191}\\
& & \overrightarrow{\nabla}\times\textbf{Q}\,=\,\textbf{B}\label{MSc192}\\
& & \overrightarrow{\nabla}\cdot\textbf{E}\,=\,\frac {\mu\varepsilon e^{*}}{D}\rho^2 \label{MSc193}\\
& & \overrightarrow{\nabla} Q+\,=\,-\textbf{E}\label{MSc194}\\
& & \frac {1}{2m^*}\Delta\rho +\alpha \rho-g\rho^3- \frac {e^{*}}{D} \rho Q-\frac {e^*}{2m^*}\rho \textbf{Q}^2=0. \label{MSc195}
\eea

There are two important differences between equations in relativistic theory (\ref{MSc111}-\ref{MSc115}) and equations in non relativistic one (\ref{MSc191}-\ref{MSc195}).
In contrast to equation (\ref{MSc113}), in equation (\ref{MSc193}) there is no dependence on gauge invariant scalar field $Q$. The last equation (\ref{MSc195}) depends on the electric potential $Q$ linearly, which makes the impact of the applied electric field on superconductivity quite nontrivial.

\section{Summary}

The aim of the present paper was to present the mathematical basis for the exploration of the intricate interplay between superconductivity and applied magnetic and electric fields. We derived  a system of equations to describe the electrodynamics of s-wave superconductors.

When the model is relativistically covariant, we obtained that the applied electric field supports the supperconductivity.
It is very difficult to experimentally test the effects of applied electric field. Right below the superconductor critical temperature the normal fluid dominates the system. The screening length of the normal fluid is about one Angstr\"{o}m. The applied field cannot penetrate into the system more than one Angstr\"{o}m from the surface. This is why the electric field cannot affect the system in the interior.  To study the effects of the applied electric field one has to do experiments at very low temperatures where there are no normal quasiparticles.

When the model is nonrelativistic, the impact of the applied electric field on superconductivity is more complicate.

The equations in relativistic theory (\ref{MSc111}-\ref{MSc115}) are invariant under the discrete transformation $\textbf{B}\rightarrow-\textbf{B}$,
$\textbf{Q}\rightarrow-\textbf{Q}$ and independently under the transformation  $\textbf{E}\rightarrow-\textbf{E}$, $Q\rightarrow-Q$. In contrast, the system of equations in nonrelativistic theory (\ref{MSc191}-\ref{MSc195}) are invariant under the discrete transformation of magnetic field $\textbf{B}$ and gauge invariant field $\textbf{Q}$, but they are not invariant under the discrete transformation of electric field $\textbf{E}$ and gauge invariant field $Q$.  This means that the effects of the applied electric fields $\textbf{E}_0$ and $-\textbf{E}_0$ on superconductivity are different.

It is important to underline that the fields $Q$ and $\textbf{Q}$ are gauge invariant. This is why they should be measurable as the electric and magnetic fields are measurable. The role of these fields is fundamental in superconductivity but not investigated.

The numerical solution of the system of equations for appropriate geometry of the superconductor, for example slab geometry \cite{FetterWalecka}, will help us to gain an insight into the impact of the applied electric field on the superconductivity.

\appendix
\section{}
To elucidate the qualitative analysis in section II, we consider the system of equations (\ref{MSc111}-\ref{MSc115}) for fields which depend on $z$ coordinate only. Then, the system of equations for the fields $Q(z)$, $\textbf{Q}(z)=(0,Q_y(z),0)$, $\textbf{E}(z)=(0,0,E_z(z))$, $\textbf{B}(z)=(B_x(z),0,0)$ and $\rho(z)$ adopts the form
\bea
& & \frac {dB_x}{dz}\,=\,-2e^{*2}\rho^2Q_y \label{MScApp1}\\
& & \frac {dQ_y}{dz}\,=\,-B_x\label{MScApp2}\\
& & \frac{dE_z}{dz}\,=\,-2e^{*2}\rho^2Q \label{MScApp3}\\
& & \frac{dQ}{dz}\,=\,-E_z\label{MScApp4}\\
& & \Delta\rho +\alpha \rho-g\rho^3+ e^{*2}\rho\left [\mu\varepsilon Q^2-Q_y^2\right]=0.  \label{MScApp5}
\eea
After some calculations one reduces the system (\ref{MScApp1}-\ref{MScApp5}) to a system of equations for $Q,Q_y$ and $\rho$
\bea
& & \frac {d^2Q}{dz^2}\,=\,2e^{*2}\rho^2Q \label{MScApp6}\\
& & \frac {d^2Q_y}{dz^2}\,=\,2e^{*2}\rho^2Q_y \label{MScApp7}\\
& & \Delta\rho +\alpha \rho-g\rho^3+ e^{*2}\rho\left [\mu\varepsilon Q^2-Q_y^2\right]=0.  \label{MScApp8}
\eea

It is convenient to introduce dimensionless functions $f_1(\zeta),f_2(\zeta)$ and $f_3(\zeta)$ of a dimensionless distance  $\zeta=z/\xi_{GL}$, where
\be\label{MScApp9}\xi_{GL}=1/\sqrt{\alpha}\ee
is the Ginzburg-Landau coherence length:
\bea\label{MScApp9}
Q(\zeta) & = & -E_0 \xi_{GL}f_1(\zeta) \nonumber \\
Q_y(\zeta) & = &  -B_0 \xi_{GL}f_2(\zeta)  \\
\rho(\zeta) & = & \rho_0 f_3(\zeta). \nonumber  \eea
In equations (\ref{MScApp9}) $\rho_0=\sqrt{\alpha/g}$, the applied electric field is $\textbf{E}_0=(0,0,E_0)$ and the applied magnetic field is
$\textbf{B}_0=(B_0,0,0)$. The representations of the electric and magnetic fields by means of $f_1$ and $f_2$ are the following:
\bea\label{MScApp10}
E_z(\zeta) & = & E_0 \frac {df_1(\zeta)}{d\zeta} \nonumber \\
B_x(\zeta) & = & B_0 \frac {df_2(\zeta)}{d\zeta}  \eea
The system of equations (\ref{MScApp6}-\ref{MScApp8}), rewritten in terms of the new functions, reads:
\bea\label{MScApp11}
& & \frac {d^2 f_1(\zeta)}{d\zeta^2}\,=\,\frac {1}{\kappa^2}f_3(\zeta)f_1(\zeta) \nonumber \\
& & \frac {d^2 f_2(\zeta)}{d\zeta^2}\,=\,\frac {1}{\kappa^2}f_3(\zeta)f_2(\zeta) \nonumber  \\
& & \frac {d^2 f_3(\zeta)}{d\zeta^2}\,+\,f_3(\zeta)\,-\,f_3^3(\zeta)\\
& & =\,-\, f_3(\zeta)\left [\gamma_E f_1^2(\zeta)-\gamma_B f_2^2(\zeta)\right]\nonumber
\eea
In equations (\ref{MScApp11}) $\kappa$ is the Ginzburg-Landau parameter
\be\label{MScApp12}
\kappa\,=\,\frac {\lambda_L}{\xi_{GL}},\ee
which satisfies $\kappa<1/\sqrt{2}$, for type I superconductors and $\kappa>1/\sqrt{2}$ for type II ones, and parameters $\gamma_E$ and $\gamma_B$ are
\be\label{MScApp13}
\gamma_E\,=\,\frac {e^{*2}\mu\varepsilon E_0^2}{\alpha^2},\hskip 1cm \gamma_B\,=\,\frac {e^{*2}B_0^2}{\alpha^2}.\ee
\begin{figure}[!ht]
\epsfxsize=\linewidth
\epsfbox{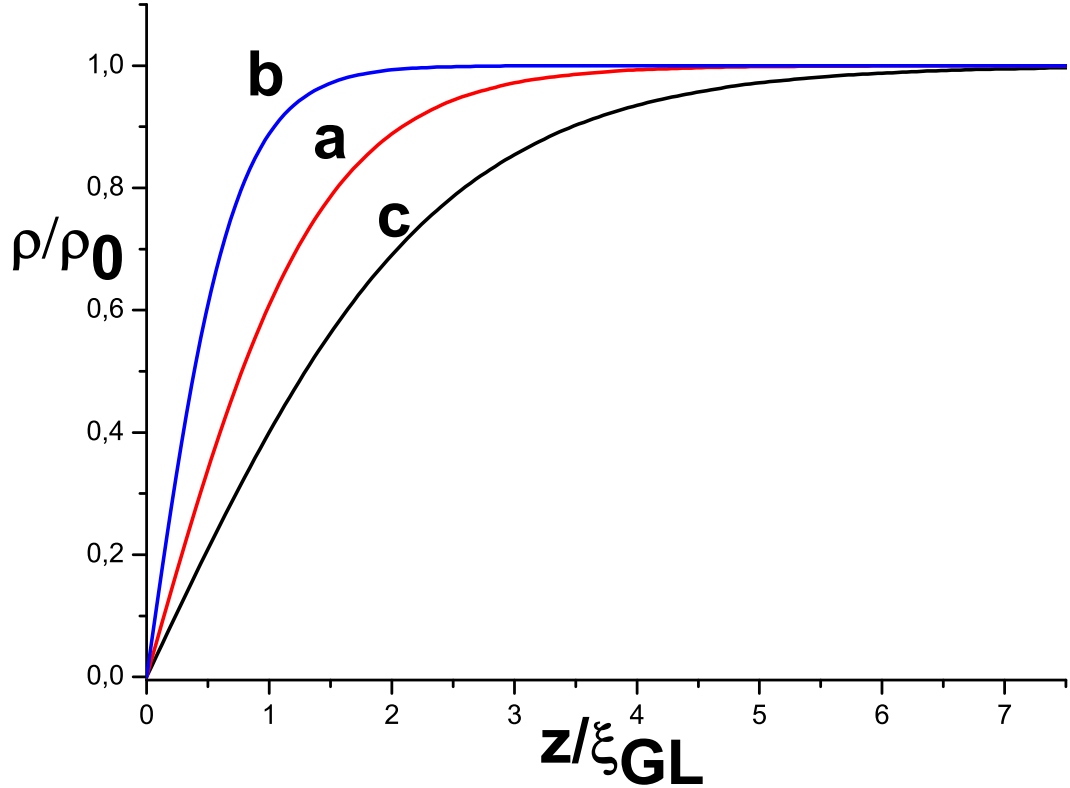} \caption{(Color online)\,\,(a)-function (\ref{MScApp15}),(b)-function (\ref{MScApp16}) $\rho(z/\xi^E)/\rho_0=\rho((\xi_{GL}/\xi^E)z/\xi_{GL})$ with $\xi_{GL}/\xi^E=2$, (c)-function (\ref{MScApp17}) $\rho(z/\xi^B)/\rho_0=\rho((\xi_{GL}/\xi^B)z/\xi_{GL})$ with $\xi_{GL}/\xi^B=0.6$.}\label{fig-MSc}
\end{figure}

For semi-infinite superconductors, with a surface of superconductor orthogonal to the $z$-axis, the boundary conditions are:
\bea\label{MScApp13}
& & \frac {df_1(0)}{d\zeta}\,=\,1 \hskip 1cm f_1(\infty)\,=\,0 \nonumber \\
& & \frac {df_2(0)}{d\zeta}\,=\,1 \hskip 1cm f_2(\infty)\,=\,0  \\
& & f_3(0)\,=\,0 \hskip 1.2cm f_3(\infty)\,=\,1 .\nonumber
\eea

If neither electric nor magnetic fields are applied the equation for the dimensionless function $f_3(\zeta)\,=\,\rho(\zeta)/\rho_0$
\be\label{MScApp14}\frac {d^2 f_3(\zeta)}{d\zeta^2}\,+\,f_3(\zeta)\,-\,f_3^3(\zeta)\,=\,0\ee
is exactly solvable  and the solution, for $z\geq 0$ is
\be\label{MScApp15}
f_3(\zeta)\,=\,f_3(\frac {z}{\xi_{GL}})\,=\,\tanh(\frac {z}{\sqrt{2}\xi_{GL}}) \ee

The qualitative analysis in section II shows that applied electric field increases the $\alpha$ parameter $\alpha\rightarrow\alpha^E=\alpha+e^{*2}\mu\varepsilon<Q^2>$, where $<Q^2>$ is an average value of the scalar field. Within this approximation the expression for $f_3^E$ is
\be\label{MScApp16}
 f_3^E(\zeta)\,=\,f_3^E(\frac {z}{\xi^E})\,=\,\tanh(\frac {z}{\sqrt{2}\xi^E}), \ee
where $\xi^E=1/\sqrt{\alpha^E}\,<\xi_{GL}$. When the magnetic field is applied $\alpha$ decreases, $\alpha\rightarrow\alpha^B=\alpha-e^{*2}<\textbf{Q}^2>$ and function $f_3^B$ reads
\be\label{MScApp17}
 f_3^B(\zeta)\,=\,f_3^B(\frac {z}{\xi^B})\,=\,\tanh(\frac {z}{\sqrt{2}\xi^B}), \ee
where $\xi^B=1/\sqrt{\alpha^B}\,>\xi_{GL}$.

The three curves (\ref{MScApp15}-\ref{MScApp17}) are depicted in figure (\ref{fig-MSc}). The graph (a) shows the function $\rho(z/\xi_{GL})/\rho_0$ (\ref{MScApp15}), when neither electric nor magnetic fields are applied, the graph (b) shows the function (\ref{MScApp16}), $\rho(z/\xi^E)/\rho_0=\rho((\xi_{GL}/\xi^E)z/\xi_{GL})$ with $\xi_{GL}/\xi^E=2$ and the graph (c) shows the function (\ref{MScApp17}) $\rho(z/\xi^B)/\rho_0=\rho((\xi_{GL}/\xi^B)z/\xi_{GL})$ with $\xi_{GL}/\xi^B=0.6$.

The Ginzburg-Landau coherence length  measures the distance over which the superconducting order parameter increases up to the bulk value, measured from the surface of the superconductor ($z>0$). The applied electric field decreases the GL coherence length, which means that the electric field supports the superconductivity, while the applied magnetic field increases the GL coherence length, which means that the magnetic field destroys the superconductivity.

\end{document}